\documentstyle[12pt,epsfig]{article}
\newcounter{minutes}
\def\lsim{\:\raisebox{-0.5ex}{$\stackrel{\textstyle<}{\sim}$}\:}

\newcommand{\newc}{\newcommand}
\newc{\pbi}{pb$^{-1}$}
\newc{\ti}{\tilde}
\newc{\ra}{\rightarrow}
\newc{\ee}{$e^+e^-$\ }
\newc{\qq}{$q\bar{q}$\ }
\newc{\dd}{$d\bar{d}$\ }
\newc{\uu}{$u\bar{u}$\ }
\newc{\eeqq}{$e^+e^-\ra q\bar{q}$\ }
\newc{\eeuu}{$e^+e^-\ra u\bar{u}$\ }
\newc{\eedd}{$e^+e^-\ra d\bar{d}$\ }
\newc{\beq}{\begin{eqnarray}}
\newc{\eeq}{\end{eqnarray}}
\newc{\dqu}{\delta_{qu}}
\newc{\dqd}{\delta_{qd}}
\newc{\non}{\nonumber}
\newc{\noi}{\noindent}

\def\ib#1,#2,#3{       {\it ibid.\/ }{\bf #1} (19#2) #3}
\def\ap#1,#2,#3{       {\it Ann.~Phys.~(NY)\/ }{\bf #1} (19#2) #3}
\def\ijmp#1,#2,#3{     {\it Int.\ J.~Mod.\ Phys.\/ } {\bf A#1} (19#2) #3}
\def\mpl#1,#2,#3 {     {\it Mod.~Phys.~Lett.\/ } {\bf A#1} (19#2) #3}
\def\npb#1,#2,#3{       {\it Nucl.\ Phys.\/ }{\bf B#1} (19#2) #3}
\def\npps#1,#2,#3{     {\it Nucl.\ Phys.~B (Proc.~Suppl.)\/ }{\bf B#1}
                             (19#2) #3}
\def\plb#1,#2,#3{      {\it Phys.\ Lett.\/ }{\bf B#1} (19#2) #3}
\def\pr#1,#2,#3{       {\it Phys.\ Rev.\/ }{\bf #1} (19#2) #3}
\def\prd#1,#2,#3{      {\it Phys.\ Rev.\/ }{\bf D#1} (19#2) #3}
\def\prep#1,#2,#3{     {\it Phys.\ Rep.\/ }{\bf #1} (19#2) #3}
\def\prl#1,#2,#3{      {\it Phys.\ Rev.\ Lett.\/ }{\bf #1} (19#2) #3}
\def\pro#1,#2,#3{      {\it Prog.~Theor.\ Phys.\/ }{\bf #1} (19#2) #3}
\def\rmp#1,#2,#3{      {\it Rev.~Mod.~Phys.\/ }{\bf #1} (19#2) #3}
\def\sp#1,#2,#3{       {\it Sov.~Phys.~Usp.\/ }{\bf #1} (19#2) #3}
\def\zpc#1,#2,#3{      {\it Z.~Phys.\/ }{\bf C#1} (19#2) #3}
\def\appb#1,#2,#3{     {\it Acta Phys.\ Polon.\/ }{\bf B#1} (19#2) #3}

\begin{document}

\begin{flushright}
BI-TP 97/07\\
DESY 97-038\\
WUE-ITP-97-02\\[1.7ex]
{\tt hep-ph/9703288} \\
\end{flushright}

\vskip 45pt
\begin{center}
{\Large \bf Leptoquark/Squark Interpretation of HERA Events: }\\[2mm]
{\Large \bf Virtual Effects in $e^+e^-$ Annihilation to Hadrons}

\vspace{11mm}
{\large J. Kalinowski}$^{1,2}$, 
{\large      R. R\"uckl}$^{3,\displaystyle \ast}$,
{\large  H. Spiesberger}$^{4,\displaystyle \ast}$,\\[1.1ex]
{\large and P.M. Zerwas}$^1$\\[2ex]
{\em $^1$ Deutsches Elektronen-Synchrotron DESY, D-22607 Hamburg}\\[1.1ex]
{\em $^2$ Institute of Theoretical Physics, Warsaw University, 
PL-00681 Warsaw}\\[1.1ex]
{\em $^3$ Institut f\"ur Theoretische Physik, Universit\"at W\"urzburg, 
D-97074 W\"urzburg}\\[1.1ex]
{\em $^4$  Fakult\"at f\"ur Physik, Universit\"at Bielefeld, 
D-33501 Bielefeld}\\[2ex]


\vspace{50pt}
{\bf ABSTRACT}
\end{center}
\begin{quotation}
  In reference to  the recently observed high $Q^2$, large $x$ events
  in deep-inelastic positron--proton scattering at HERA, various 
  leptoquark and supersymmetric scenarios are discussed. 
  We study the impact of virtual leptoquark or $R$-parity breaking
  squark exchange as well as generic contact interaction 
  on the production of quark--antiquark pairs in \ee
  annihilation, in particular at LEP2. 
\end{quotation}

\vspace*{\fill}
\footnoterule
{\footnotesize
\noindent ${}^{\displaystyle \ast}$
Supported by Bundesministerium f\"ur Bildung,
Wissenschaft, Forschung und Technologie, Bonn, Germany, Contracts
05 7BI92P (9) and 05 7WZ91P (0).}

\newpage
\renewcommand{\thefootnote}{\arabic{footnote}}

\section{Introduction}
The recent observation of events in deep-inelastic positron--proton
scattering with very high $Q^2$ and large $x$ at HERA \cite{sem} has
refuelled speculations on physics beyond the Standard Model, in
particular on low-mass leptoquark-type particles. Such particles 
had been suggested
a long time ago in a variety of physical scenarios: Pati--Salam SU(4)
unification of quarks and leptons \cite{PS}, grand unified theories
such as SU(5) or E$_6$ \cite{GG}, and composite models
\cite{composite}.  Moreover, in supersymmetric theories squarks couple
to lepton--quark pairs if the $R$-symmetry is broken in the trilinear
couplings of the superfields \cite{Rbroken,suprp}. Vector leptoquarks in
grand unified theories with both lepton-quark and diquark couplings
 must be very heavy to suppress proton decay; certain  
scalar leptoquarks in GUT multiplets could nevertheless be relatively light
\cite{yanagida} (disregarding the notorious  hierarchy problem
for the time being).  Squarks in supersymmetric theories should
naturally be expected in the mass range of a few hundred GeV.

A general classification of these novel states\footnote{We shall
  generically denote leptoquarks and squarks in $R$-parity breaking
  scenarios by $LQ$.} has been presented in Ref.\ \cite{BRW}. In this
analysis the couplings of leptoquarks to lepton--quark pairs are
assumed to be baryon- and lepton-number conserving in order to avoid
rapid proton decay, family diagonal to exclude FCNC processes beyond the 
CKM mixing,
and chiral to preserve the helicity suppression in leptonic pion
decay. Moreover, the couplings are taken dimensionless and
all interactions are assumed to respect the
SU(3)$_C\times$SU(2)$_L\times$U(1)$_Y$ symmetry of the Standard Model.
The allowed states can be classified according to spin, weak isospin
and fermion number. We adopt the notation of Refs.\ \cite{Aachen} to
conform with the notation generally employed in experimental papers:
vector leptoquarks are denoted by $V_I$, scalar leptoquarks by $S_I$; 
isomultiplets with different hypercharges are distinguished by a tilde.

{\small
\begin{table}[htp]
\begin{center}
\begin{tabular}{|c|c|c|c|c|c||c|c||c|}
\hline
\multicolumn{2}{|c|}{\rule{0mm}{5mm}$LQ$} & 
  $Q$ & 
     Decay & BR &
    Coupling & 
   Limits & HERA &
      $e^-_ie^+\ra q_k\bar{q}$
       \\
\multicolumn{2}{|c|}{}&
  &
     Mode & $e^{\pm}\,j$ &
       & Ref.\ \cite{Leurer,Davidson} & estimates&
      $ik$
        \\[1mm]
%
\hline \rule{0mm}{5mm}
$S_0$ & {\raisebox{1.ex}{$\displaystyle
        \begin{array}{c}\rule{0mm}{5mm}\tilde{d}_R\\ ~ \end{array}$}} &
  $-1/3 $ & 
    $\displaystyle \begin{array}{c} e_L u  \\ \nu_L d\\ e_R u \end{array}$ & 
     {\raisebox{-0.8ex}{$\displaystyle 
     \begin{array}{c} \frac{1}{2} \\ \rule{0mm}{6mm} 1 \end{array}$}} &
    $\displaystyle \begin{array}{c} g_L \\ -g_L\\ g_R  \end{array}$ &  
     {\raisebox{-0.8ex}{$\displaystyle 
     \begin{array}{c} g_L<0.06 \\ \rule{0mm}{6mm} g_R<0.1 \end{array}$}} &
     {\raisebox{-0.8ex}{$\displaystyle 
     \begin{array}{c} 0.40 \\ \rule{0mm}{6mm} 0.28 \end{array}$}} &
      $\displaystyle \begin{array}{c} LL  \\ - \\RR\end{array}$   
                     \\[1mm]
%
\hline
\multicolumn{2}{|c|}{\rule{0mm}{5mm}$\tilde{S}_0$} & 
  $-4/3 $      &
     $e_R d$ & 1 &
    $g_R$ &
   $g_R<0.1$    & 0.30 &
      $RR$
        \\[1mm]
%
\hline
\multicolumn{2}{|c|}{\rule{0mm}{5mm}} &  
  $+2/3$ & 
     $\nu_L u$ & 0 &
    $\sqrt{2}g_L$  & 
                & $-$ & 
      $-$ 
        \\
\cline{3-6}\cline{9-9}
\multicolumn{2}{|c|}{$S_1$} &
  $-1/3$  & 
     $\displaystyle \begin{array}{c} \nu_L d \\ e_L u \end{array}$ &  
           $\frac{1}{2}$ &
    $\displaystyle \begin{array}{c} -g_L \\ -g_L \end{array}$ & 
   $g_L<0.09$    & 0.40 &
      $\displaystyle \begin{array}{c} - \\ LL  \end{array}$
        \\
\cline{3-6}\cline{9-9}
\multicolumn{2}{|c|}{\rule{0mm}{5mm}} & 
$-4/3$ 
  & 
     $e_L d$ & 1 & 
    $-\sqrt{2}g_L$ & 
                 & 0.21 & 
      $LL$
        \\[1mm]
%
\hline
\multicolumn{2}{|c|}{\rule{0mm}{5mm}} &  
  $-1/3$ &
     $\displaystyle \begin{array}{c} \nu_L d \\ e_R u \end{array}$ & 
     $\displaystyle \begin{array}{c} 0 \\ 1 \end{array}$ & 
    $\displaystyle \begin{array}{c} g_L \\ g_R \end{array}$&
   $g_L<0.09$& 
    $\displaystyle \begin{array}{c} - \\ 0.30 \end{array}$ &
      $\displaystyle \begin{array}{c} - \\ RL \end{array}$
        \\[1mm] 
\cline{3-6}\cline{9-9}
\multicolumn{2}{|c|}{{\raisebox{ 3ex}[-3ex]{ $V_{1/2}$}}}& 
  $-4/3$  & 
     $\displaystyle \begin{array}{c} e_L d  \\  e_R d \end{array}$ &
        $1$ &
    $\displaystyle \begin{array}{c} g_L \\ g_R\end{array}$&
   $g_R<0.05$& 
    $\displaystyle \begin{array}{c} 0.32 \\ 0.32 \end{array}$ &
      $\displaystyle \begin{array}{c} LR \\ RL\end{array}$
        \\ [1mm]
%
\hline
\multicolumn{2}{|c|}{\rule{0mm}{5mm}} &   
  $ +2/3$ & 
     $\nu_L u$& 0 & 
    $g_L$ &
            & $-$ &
      $-$
        \\[1mm]
\cline{3-6}\cline{9-9}
\multicolumn{2}{|c|}{{\raisebox{ 1.5ex}[-1.5ex]{ $\tilde{V}_{1/2}$}}}&
\rule{0mm}{5mm}
  $ -1/3$ & 
     $ e_L u $& 1&
    $g_L$ &
   {\raisebox{ 1.6ex}[-1.6ex]{$g_L<0.09$}} & 0.32 & 
      $LR$ 
        \\[1mm]
\hline
\hline
\multicolumn{2}{|c|}{\rule{0mm}{5mm}} &  
  $-2/3$ &
     $\displaystyle\begin{array}{c}\nu_L \bar{u}\\e_R \bar{d} \end{array}$ &
     $\displaystyle\begin{array}{c} 0 \\ 1 \end{array}$ &
    $\displaystyle\begin{array}{c} g_L \\ -g_R \end{array}$&
   $g_L<0.1$& 
    $\displaystyle\begin{array}{c} - \\ 0.052 \end{array}$ &
      $\displaystyle \begin{array}{c}-\\ RL \end{array}$
        \\[1mm] 
\cline{3-6}\cline{9-9}
\multicolumn{2}{|c|}{{\raisebox{ 3ex}[-3ex]{ $S_{1/2}$}}}& 
  $-5/3$  & 
     $\displaystyle\begin{array}{c} e_L \bar{u}\\ e_R \bar{u}\end{array}$&
        1 &
    $\displaystyle\begin{array}{c} g_L \\ g_R\end{array}$&
   $g_R<0.09$& 
    $\displaystyle\begin{array}{c} 0.026 \\ 0.026 \end{array}$ &
      $\displaystyle \begin{array}{c} LR \\ RL\end{array}$
        \\ [1mm]
%
\hline
\rule{0mm}{6mm} & $\overline{\tilde{d}}_L$ &  
  $ +1/3$ & 
     $\nu_L \bar{d}$& 0 & 
    $g_L$ &
             & $-$ & 
      $-$
        \\[1mm]
\cline{2-6}\cline{9-9}
\rule{0mm}{5mm}
{\raisebox{2ex}[-2ex]{ $\begin{array}{c} \tilde{S}_{1/2} 
                     \\ \ast \end{array}$}} & 
$\overline{\tilde{u}}_L$&
  $ -2/3$ &
     $ e_L\bar{d} $& 1 &
    $g_L$ &
   {\raisebox{2.5ex}[-2.5ex]{$g_L<0.1$}}& 0.052 &
      $LR$
         \\[1mm]
%
%
\hline
\multicolumn{2}{|c|}{\rule{0mm}{6mm}$V_0$} & 
  $-2/3 $ & 
     $\displaystyle \begin{array}{c} e_L \bar{d}  \\
                     \nu_L \bar{u}\\ e_R \bar{d} \end{array}$ & 
     {\raisebox{-0.8ex}{$\displaystyle 
     \begin{array}{c} \frac{1}{2} \\ \rule{0mm}{6mm} 1 \end{array}$}} &
    $\displaystyle \begin{array}{c} g_L \\ g_L\\ g_R  \end{array}$ &  
     {\raisebox{-0.8ex}{$\displaystyle 
     \begin{array}{c} g_L<0.05 \\ \rule{0mm}{6mm} g_R<0.09 \end{array}$}} &
     {\raisebox{-0.8ex}{$\displaystyle 
     \begin{array}{c} 0.080 \\ \rule{0mm}{6mm} 0.056 \end{array}$}} &
      $\displaystyle \begin{array}{c} LL  \\ - \\RR\end{array}$
                     \\[1mm]
%
\hline
\multicolumn{2}{|c|}{\rule{0mm}{5mm}$\tilde{V}_0$} & 
  $-5/3$      & 
     $e_R \bar{u}$ & 1 &
    $g_R$   &
   $g_R<0.09$& 0.027 &
      $RR$
        \\[1mm]
%
\hline
\multicolumn{2}{|c|}{\rule{0mm}{5mm}} &  
  $+1/3$         & 
     $\nu_L \bar{d}$ & 0 &
    $\sqrt{2}g_L$  & 
                     & $-$ &
      $-$ 
        \\
\cline{3-6}\cline{9-9}
\multicolumn{2}{|c|}{$V_1$} &
  $-2/3$&
     $\displaystyle\begin{array}{c} e_L \bar{d}\\ \nu_L\bar{u}\end{array}$&
         $\frac{1}{2}$ &
    $\displaystyle\begin{array}{c}\rule{0mm}{5mm} -g_L\\g_L \end{array}$ & 
   $g_L<0.04$& 0.080& 
      $\displaystyle \begin{array}{c} LL\\- \end{array}$
        \\
\cline{3-6}\cline{9-9}
\multicolumn{2}{|c|}{\rule{0mm}{5mm}} &  
  $-5/3$ & 
     $e_L \bar{u}$ & 1 &
    $\sqrt{2}g_L$ & 
                    & 0.019 & 
      $LL$ 
        \\[1mm]
\hline
\end{tabular}
\caption{\it Scalar ($S$) and vector ($V$) leptoquarks/squarks with  
  electric charges ($Q$), decay modes,
  branching ratios for charged lepton + jet channels with either 
$L$ or $R$ couplings, and the Yukawa
  couplings ($g_{R,L}$) with the most stringent limits from rare decays 
  and estimates from the recent 
  HERA data (see text).  The helicity combinations $ik$ ($=L,R$)
  contributing to the process $e^-_i e^+\ra q_k\bar{q}$ are given in
  the last column.  Also shown are possible squark assignments of
  leptoquark-type states;  the special 
   leptoquark singled out in Ref.\ \protect\cite{yanagida} is
  marked by an asterisk.}
\label{tabprop}
\end{center}
\end{table}}

For convenience the nine possible states of scalar and vector
leptoquarks are listed in Table \ref{tabprop}.  Leptoquarks in the
upper (lower) part of Table \ref{tabprop} carry fermion number $F=2$
($F=0$). The couplings are denoted generically by $g_{R}$ or $g_L$
with $R,L$ refering to the chirality of the lepton.  Each state can
couple with different strength; for simplicity, the additional indices
are suppressed.  In principle the two scalar states $S_0$ and
$S_{1/2}$ and the two vector states $V_0$ and $V_{1/2}$ could have
both chiral $g_R$ and $g_L$ couplings at the same time; however, since
the product of the two couplings is constrained very strongly by rare
decays~\cite{BW,Leurer,Davidson,beta}, we assume only one of the two
couplings to be non-zero. The special type of leptoquark that does not
induce proton decay (as a result of its quantum numbers) and is
compatible with the renormalization of the electroweak mixing angle
from the symmetry value 3/8 at the GUT scale down to
$\sin^2\Theta_W\cong 0.23$ at the electroweak scale~\cite
{yanagida}, is marked by an asterisk.  Only a small subset of all
these states is realized in supersymmetric theories with $R$-parity
breaking.  Moreover, supersymmetry requires these states to have
universal left-handed couplings to leptons.

If leptoquarks or squarks in $R$-parity breaking supersymmetric
theories exist, a large variety of phenomena are expected to be
observed experimentally.  In electron/positron--proton collisions,
these particles are produced as single resonances, with the rate
determined by the strength of the $LQ$-$l$-$q$ Yukawa couplings
\cite{BRW,butt}.  Pair production is an important production mechanism for
leptoquarks in proton--(anti)proton \cite{pp}, electron--positron
\cite{ee} and photon--photon collisions \cite{gamgam}. In these
reactions, the size of the cross section is determined ({\it modulo}
anomalous couplings and form-factor effects) by the color, the
electric and the electroweak charges of a given leptoquark (if the
Yukawa couplings are small \cite{BR}).  The predictions for the cross
sections are therefore, to a first approximation, model-independent.
Associated production of $LQ$+$l$ or $LQ$+$q$, again 
mediated by Yukawa interactions, can also be explored in these and
other collision processes ($e\gamma$, for example~\cite{who}).
 
In addition to the direct production of leptoquarks/squarks, which is of
course of central interest, indirect effects generated by the exchange
of virtual $LQ$s, are also important experimental tools to provide
cross checks, to explore the nature of these particles, and to have a
glimpse at states which are too heavy to be produced directly.
Virtual leptoquarks could strongly affect decay processes such as
$\pi\ra l\nu$ and $K\ra\pi\nu\bar{\nu}$. These processes are
suppressed for the $LQ$s considered here by the restrictions on the
Yukawa couplings summarized above. Nevertheless, they still contribute
to rare pion and kaon decays, $K^0-\bar{K}^0$ and $D^0-\bar{D}^0$
mixing, and atomic parity violation \cite{BW,Leurer,Davidson}.  The
strongest constraints can be deduced from atomic parity violation with
two exceptions: the constraints on $g_{L}(S_0)$ and $g_{L}(V_0)$
which follow from the violation of universality in $\pi \rightarrow l
\nu$ \cite{Leurer} or in $\mu$ and $\beta$ decays
\cite{Davidson,beta}. The bounds, as derived in Refs.\ \cite{Leurer}
and \cite{Davidson}, are presented in Table \ref{tabprop} for the
reference mass $m_{LQ}$ = 200 GeV; the leptoquarks are taken mass
degenerate within the isospin multiplets. In the few hundred GeV range
the bounds on the couplings scale linearly with the leptoquark mass.
Potential destructive interference effects between different states
can lift these bounds considerably.

Additional indirect constraints can be derived from the $t/u$ channel exchange
of lepto\-quarks/\-squarks in high energy processes, such as Drell--Yan
production of lepton pairs in $pp$ and $p\bar{p}$ collisions
\cite{pp2}, or \ee annihilation to hadrons \cite{ee2}.  The existing
bounds can be improved significantly with the rise of the \ee energy
to the  LEP2  value close to $\sqrt{s}=200$ GeV.

\begin{figure}[htb]
\centerline{\psfig{figure=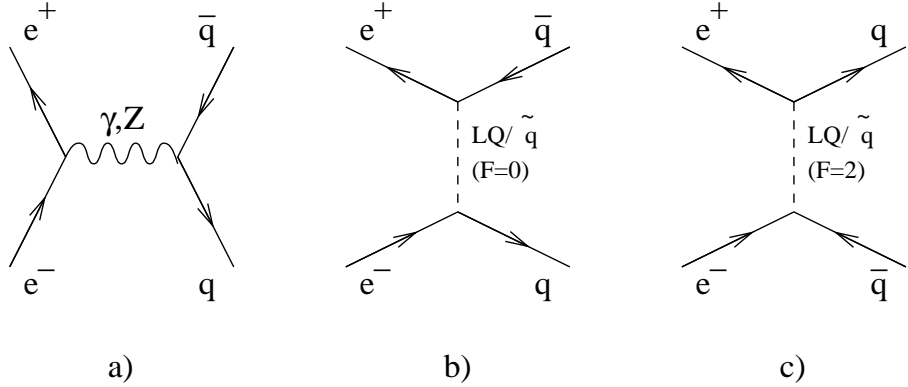,height=12cm,width=5cm,angle=-90}}
\caption{\it Feynman diagrams for \eeqq including leptoquarks and squarks
in $R$-parity breaking supersymmetric theories.}
\label{diag}
\end{figure}

In the present analysis we study the production of \qq pairs
in \ee annihilation, \beq e^+e^- \ra q\bar{q} \label{process} \eeq
which is mediated by $\gamma,Z$ exchanges in the $s$-channel (Fig.\ 
\ref{diag}a), and leptoquark/squark exchanges in the $t/u$-channels
(Fig.\ \ref{diag}b and c).  By definition, $t$-channel exchange is
associated with $e^-\ra q$ transitions ($F=0$), while $u$-channel
exchange involves $e^-\ra\bar{q}$ transitions ($F=2$). We will
consider both scalar and vector leptoquarks. Earlier analyses of Ref.\ 
\cite{ee2} will be extended in several aspects: (i) We present a
systematic analysis of the exchange of all types of leptoquarks. In
particular, generalizing the helicity method of Ref.\ \cite{SZ}, a
representation of the exchange amplitudes can be derived in which
potential interference effects between different types of leptoquarks
are made transparent. (ii) Simple expressions for the integrated cross
sections are presented. (iii) Special emphasis is given to squark
exchange mechanisms in supersymmetric theories with $R$-parity
breaking, complementing earlier discussions in Ref.\ \cite{debchou}.

\section{Leptoquark/Squark Exchange in \ee Annihilation}
After performing the standard Fierz transformations from
$t/u$-channel leptoquark/squark exchange amplitudes in \eeqq 
to  standard $s$-channel amplitudes, only
terms of the structure {\it (lepton vector
current)}$\times${\it (quark vector current)} are generated for 
leptoquarks carrying either left {\it or} right chiral couplings, 
but not {\it both} at the same time. This leads
to a  transparent representation of the matrix elements. 
Denoting the helicity amplitudes of the process $e^-_ie^+\ra
q_k\bar{q}$ with $i,k=R,L$ by $f_{ik}$ and the spin-density matrix
elements, which depend on the polar angle $\theta$ between $e^-$ and
$q$, by $\rho_{ik}$, the cross section for the process
Eq.\ (\ref{process}) can be written as
\beq
\frac{\mbox{d}\sigma}{\mbox{d}\cos\theta}(e^+e^-\ra q\bar{q}) =
\frac{N_c}{128\pi s} \sum_{i,k=L,R} \rho_{ik} |f_{ik}|^2 \label{diff-xsec} 
\eeq
with the color factor $N_c=3$.
As a consequence of angular momentum conservation, the spin-density 
matrix elements are given by 
\beq 
&\rho_{RR}=\rho_{LL}=s^2(1+\cos\theta)^2\\ 
&\rho_{RL}=\rho_{LR}=s^2(1-\cos\theta)^2 \eeq 
The helicity amplitudes can be cast
in the following form {\small 
\beq 
f_{RR}&=&\frac{Q^{eq}_{RR}}{s}
+\dqu\left(\frac{g^2_{R}}{t-\ti{V}_0} -\frac{g^2_{R}}{2(u-S_0)}\right)
+\dqd\left(\frac{g^2_{R}}{t-V_0} -\frac{g^2_{R}}{2(u-\ti{S}_0)}\right)
\non \\ 
f_{LL}&=&\frac{Q^{eq}_{LL}}{s}
+\dqu\left(\frac{2g^2_{L}}{t-V_1} -\frac{g^2_{L}}{2(u-S_0)}
-\frac{g_{L}^2}{2(u-S_1)}\right)
+\dqd\left(\frac{g^2_{L}}{t-V_0} +\frac{g^2_{L}}{t-V_1}
-\frac{g^2_{L}}{u-{S}_1}\right) \non \\ 
f_{RL}&=&\frac{Q^{eq}_{RL}}{s}
+\dqu\left(\frac{g^2_{R}}{2(t-S_{1/2})}
-\frac{g^2_{R}}{u-V_{1/2}}\right)
+\dqd\left(\frac{g^2_{R}}{2(t-S_{1/2})}
-\frac{g^2_{R}}{u-V_{1/2}}\right) \non \\ 
f_{LR}&=&\frac{Q^{eq}_{LR}}{s} +\dqu\left(\frac{g^2_{L}}{2(t-S_{1/2})}
-\frac{g^2_{L}}{u-\ti{V}_{1/2}}\right)
+\dqd\left(\frac{g^2_{L}}{2(t-\ti{S}_{1/2})}
-\frac{g^2_{L}}{u-V_{1/2}}\right)
\label{hel} \eeq   }
\noindent
To simplify the notation in the above formulae, we have used the 
$LQ$ symbols to denote the mass squared, $LQ:=m^2_{LQ}$;  
the couplings are  denoted generically by $g_R$ or $g_L$ with indices
identifying the leptoquark type suppressed. The generalized charges 
in the  standard $\gamma,Z$ exchange amplitudes have been abbreviated by 
$Q^{eq}_{ik}$ where 
\beq Q^{eq}_{ik}= e^2\,Q_eQ_q +\frac{g^e_ig^q_k}{1-m_Z^2/s} 
\eeq
with the left/right $Z$ charges of the fermions  defined as 
\beq g^f_L&=&\frac{e}{s_Wc_W}\left[I^f_3-s^2_W Q_f\right] \non\\
     g^f_R&=&\frac{e}{s_Wc_W}\left[ {} -s^2_W Q_f\right]\non
\eeq
and $s_W=\sin\Theta_W$, $c_W=\cos\Theta_W$.  The Mandelstam variables
$t,u$ can be expressed by the production angle $\theta$:
$t=-s(1-\cos\theta)/2$, $u=-s(1+\cos\theta)/2$; they are both negative
so that the amplitudes for $LQ$ exchange do not change the sign when
$\theta$ is varied from the forward to the backward direction.

It is obvious from the expressions (\ref{hel}) that 
leptoquarks/squarks of a
given fermion number $F$ contribute with a fixed positive or negative
sign to the helicity amplitude $f_{ik}$, thus reinforcing their impact
mutually.  For a given $F$, the sign of the interference with
$\gamma/Z$ exchange is determined by the sign of the generalized
charges $Q_{ik}^{eq}$, which, in the energy range considered here, are
negative for $u$-quarks and positive for $d$-quarks, except
$Q_{RL}^{ed}$ which is negative.  By contrast, leptoquarks with
different fermion numbers interfere destructively with each other.
Leptoquarks with integer isospin contribute to equal-helicity $LL$ and
$RR$ amplitudes, while leptoquarks with $I=1/2$ contribute to
opposite-helicity amplitudes $RL$ and $LR$.

{\small
\newcommand{\ff}{\frac{1}{4}}
\begin{table}[ht]
\begin{center}
\begin{tabular}{|c||c|c|c|c||c|c|c|c|}
\hline
\rule{0mm}{5mm} 
&\multicolumn{4}{c||}{\uu final state}&\multicolumn{4}{c|}{\dd final state}\\
\cline{2-9} \rule{0mm}{5mm}
$LQ$   &    $k_1$   &    $k_2$   &   $k_3$   &   $k_4$  
  &  $k_1$   &    $k_2$   &   $k_3$   & $k_4$  \\
[1mm]\hline \hline \rule{0mm}{5mm} 
$S_0$ &     &   $-Q^{eu}_{RR}g^2_{R}$ & & $\ff g^4_{R}$ 
&&&&\\
\rule{0mm}{5mm} 
$S_0$ &     &   $-Q^{eu}_{LL}g^2_{L}$ & & $\ff g^4_{L}$ 
 &&&& \\
[1mm]\hline \rule{0mm}{5mm} 
$\ti{S}_0$    & &                        & &           
 &  & $ -Q^{ed}_{RR}g^2_{R}$ & & $\ff g^4_{R}$ \\
[1mm]\hline \rule{0mm}{5mm} 
$S_1$  &     &   $-Q^{eu}_{LL}g^2_{L}$ & & $ \ff g^4_{L}$ 
 &     &   $-2Q^{ed}_{LL}g^2_{L}$ & & $g^4_{L}$\\
[1mm]\hline \rule{0mm}{5mm}
$V_{1/2}$   &&                       & &           
 &     $-2Q^{ed}_{RL}g^2_{R}$ & & $g^4_{R}$   &   \\
$V_{1/2}$     & $-2Q^{eu}_{RL}g^2_{R}$ & & $ g^4_{R}$  & 
 &     $-2Q^{ed}_{LR}g^2_{L}$ & & $g^4_{L}$   &\\
[1mm]\hline \rule{0mm}{5mm} 
$\ti{V}_{1/2}$& $-2Q^{eu}_{LR}g^2_{L}$ & & $ g^4_{L}$ & 
 &&&&\\
[1mm]\hline\hline
\rule{0mm}{5mm} 
$S_{1/2}$    &&  $Q^{eu}_{RL}g^2_{R}$ & & $ \ff g^4_{R}$ 
 &  &    $Q^{ed}_{RL}g^2_{R}$ & & $\ff g^4_{R}$ \\
\rule{0mm}{5mm} 
$S_{1/2}$   &&  $Q^{eu}_{LR}g^2_{L}$ & & $\ff g^4_{L}$
 &&&& \\
[1mm]\hline \rule{0mm}{5mm}
$\ti{S}_{1/2}$       &&                       & &              
 && $Q^{ed}_{LR}g^2_{L}$ & & $\ff g^4_{L}$\\
[1mm]\hline \rule{0mm}{5mm}
$V_0$ &&&&
 &         $ 2Q^{ed}_{RR}g^2_{R}$ & & $g^4_{R}$   &\\
\rule{0mm}{5mm} 
$V_0$ &&&&
&         $ 2Q^{ed}_{LL}g^2_{L}$ & & $g^4_{L}$ &\\
[1mm]\hline \rule{0mm}{5mm} 
$\ti{V}_0$  &    $2Q^{eu}_{RR}g^2_{R}$ & & $ g^4_{R}$ & 
&&&&\\
[1mm]\hline \rule{0mm}{5mm} 
$V_1$  &        $ 4Q^{eu}_{LL}g^2_{L}$ & & $4g^4_{L}$ &
 &         $ 2Q^{ed}_{LL}g^2_{L}$ & & $g^4_{L}$   & \\
[1mm]\hline \end{tabular}
\caption{\it The coefficients $k_i$ in  Eq.~(\ref{total}) describing 
the exchange of leptoquarks in the total cross section of \ee 
annihilation to hadrons.} \label{tabcoeff}
\end{center}
\end{table}}

The angular integration can easily be performed to obtain 
the total cross section for 
$e^+e^-\ra q\bar{q}$  including the exchange of one leptoquark 
with either left or right coupling: 
\beq
\sigma(e^+e^-\ra q\bar{q})=\frac{N_c}{128\pi s}\,
\left[\, \frac{8}{3}\, 
\left(|Q^{eq}_{RR}|^2+|Q^{eq}_{LL}|^2+|Q^{eq}_{RL}|^2+|Q^{eq}_{LR}|^2
\right) + \sum^4_{i=1} k_i\, C_i(\tau) \right] \label{total}
\eeq
The ratio $\tau$ is defined as $\tau=m_{LQ}^2/s$.  
If two or more
leptoquarks contribute to the same helicity amplitude, interference
terms between pairs of leptoquarks must be included; they are
collected in the Appendix. The interference terms between leptoquark
and $\gamma,Z$ exchange amplitudes are described by two functions,
depending on the mass ratio $\tau$,
\beq
C_1(\tau)&=&  12+8\tau-8(1+\tau)^2\log\frac{1+\tau}{\tau}\\
C_2(\tau)&=&  8\tau-4-8\tau^2\log\frac{1+\tau}{\tau}
\eeq
The couplings building up $k_1$ and $k_2$ are listed in Table 
\ref{tabcoeff}.  The squared leptoquark-exchange 
amplitudes are  given by two additional  functions,
\beq
C_3(\tau)& =&16+\frac{8}{\tau}-16(1+\tau)\log\frac{1+\tau}{\tau}\\
C_4(\tau)&=&16-\frac{8}{1+\tau}-16\tau\log\frac{1+\tau}{\tau}
\eeq
with the  coefficients  $k_3$ and $k_4$ again listed in  
Table~\ref{tabcoeff}.

It is instructive to consider the helicity amplitudes explicitly in
the large mass limit $m_{LQ}\gg\sqrt{s}$. In this limit they can be
interpreted as lepton-quark contact terms:
\beq 
f_{RR}&=&\frac{Q^{eq}_{RR}}{s}
-\dqu\left(\frac{g^2_{R}}{\ti{V}_0} -\frac{g^2_{R}}{2S_0}\right)
-\dqd\left(\frac{g^2_{R}}{V_0} -\frac{g^2_{R}}{2\ti{S}_0}\right)
\non \\ 
f_{LL}&=&\frac{Q^{eq}_{LL}}{s}
-\dqu\left(\frac{2g^2_{L}}{V_1} -\frac{g^2_{L}}{2S_0}
-\frac{g_{L}^2}{2S_1}\right)\non 
-\dqd\left(\frac{g^2_{L}}{V_0} +\frac{g^2_{L}}{V_1}
-\frac{g^2_{L}}{{S}_1}\right) \non \\ 
f_{RL}&=&\frac{Q^{eq}_{RL}}{s}
-\dqu\left(\frac{g^2_{R}}{2S_{1/2}}
-\frac{g^2_{R}}{V_{1/2}}\right)
-\dqd\left(\frac{g^2_{R}}{2S_{1/2}}
-\frac{g^2_{R}}{V_{1/2}}\right) \non \\ 
f_{LR}&=&\frac{Q^{eq}_{LR}}{s} -\dqu\left(\frac{g^2_{L}}{2S_{1/2}}
-\frac{g^2_{L}}{\ti{V}_{1/2}}\right)
-\dqd\left(\frac{g^2_{L}}{2\ti{S}_{1/2}}
-\frac{g^2_{L}}{V_{1/2}}\right)
\label{helcont} \eeq
As observed before, leptoquarks with integer isospin $I=0$ and $I=1$ 
build up equal-helicity $LL$ and $RR$ contact terms, while leptoquarks with
$I=1/2$ contribute to opposite-helicity  $RL$ and $LR$ contact terms.
These rules may also be cast into the standard form of the 
effective Lagrangian~\cite{effl}
\beq
{\cal L}_{eff}&=&\sum_{i,k=L,R}\frac{ g^2_i}{m^2_{LQ}}\; \alpha^{ik}\;
(\bar{e}_i\gamma^{\mu} e_i)\,(\bar{q}_k\gamma_{\mu} q_k)
\non\\ &:=& 
\sum_{i,k=L,R}\eta_{ik}\;\frac{ 4\pi}{\Lambda^2_{ik}}\;
(\bar{e}_i\gamma^{\mu} e_i)\,(\bar{q}_k\gamma_{\mu} q_k)
\label{Leff} \eeq
 
\newcommand{\ft}{$\frac{1}{2}$}
\newcommand{\fm}{$-\frac{1}{2}$}
\newcommand{\lin}{\\[1mm]\hline \rule{0mm}{5mm}} 
{\small 
\begin{table}[ht]
\begin{center}
\begin{tabular}{|c||c|c|c|c||c|c|c|c|}
\hline
\rule{0mm}{5mm} 
&\multicolumn{4}{c||}{\uu final state}&\multicolumn{4}{c|}{\dd final state}\\
\cline{2-9} \rule{0mm}{5mm}
$\alpha^{ik}$          &$RR$ &$LL$ &$RL$ &$LR$ &$RR$ &$LL$ &$RL$ &$LR$ \\
[1mm]\hline \hline \rule{0mm}{5mm} 
$S_0$         & \ft & \ft &     &     &     &     &     &\lin
$\ti{S}_0$    &     &     &     &     &\ft  &     &     & \lin 
$S_1$         &     & \ft &     &     &     & $1$ &     & \lin 
$V_{1/2}$     &     &     & $1$ &     &     &     & $1$ & $1$  \lin
$\ti{V}_{1/2}$&     &     &     & $1$ &     &     &     & \\
[1mm]\hline \hline \rule{0mm}{5mm} 
$S_{1/2}$     &     &     & \fm & \fm &     &     & \fm & \lin
$\ti{S}_{1/2}$&     &     &     &     &     &     &     & \fm \lin
$V_0$         &     &     &     &     &$-1$ & $-1$&     & \lin
$\ti{V}_0$    &$-1$ &     &     &     &     &     &     & \lin
$V_1$         &     &$-2$ &     &     &     &$-1$ &     & \\
[1mm]\hline \end{tabular}
\caption{\it The coefficients $\alpha^{ik}$ in the Lagrangian of the 
contact interactions.}\label{tabcon1}
\end{center}
\end{table}}

\noindent
with $e_i$, $q_k$ denoting left- and right-handed electron and quark 
fields.  The
coefficients $\alpha^{ik}$ for \uu and \dd final states are listed in
Table~\ref{tabcon1}. Denoting the signs of $\alpha^{ik}$ by $\eta_{ik}$, 
the scales $\Lambda_{ik}^2$ 
of the contact interactions are related to the individual masses and 
couplings of the leptoquarks by $\Lambda^2_{ik}=4\pi m^2_{LQ}/g^2_i 
|\alpha^{ik}|$.
In the total cross section $\sigma(e^+e^-\ra q\bar{q})$ the 
interference  terms and the squared  contact terms 
approach the  limits $C_1=C_2=-8s/3m_{LQ}^2$ and $C_3=C_4=8s^2/3m_{LQ}^4$ 
for $m_{LQ}\gg\sqrt{s}$.

\section{Phenomenological Evaluation}
If the surplus of the HERA high $Q^2$, large $x$ events is interpreted
as the production of scalar or vector leptoquarks, or of squarks, 
their Yukawa
couplings can be estimated from the production rates. We present only
{\it qualitative estimates} of these couplings which should illustrate
the general expectations for possible effects in \ee annihilation but 
which should not
anticipate a rigorous analysis to be performed by the experiments
themselves. Nevertheless, averaging over the H1 and ZEUS data 
one finds the couplings listed in Table~\ref{tabprop}.  
It is assumed in these estimates that only one type of
leptoquark ($i.e.$ a single member of an isomultiplet) 
has been generated with one specific chiral 
coupling ($L,R$)  to one specific quark flavor  
($up$, $down$) 
which gives rise to the branching ratios for the decays
of leptoquarks into charged leptons shown in the same Table. The
$F=2$ leptoquarks (upper part of the Table) are generated in positron 
sea-quark collisions, the $F=0$ leptoquarks (lower part of the Table) 
in positron valence-quark collisions.

From the couplings shown in Table \ref{tabprop} we can draw interesting
conclusions even at the present level of qualitative arguments: 
\\[2mm] 
(i) If the HERA events are interpreted as the signal of a
leptoquark generated in positron valence-quark collisions, the Yukawa
coupling is of the order $\sim e/10$, $i.e.$ suppressed by a full
order of magnitude compared with the electromagnetic coupling. The
$t/u$-channel exchange of such a state affects the \eeuu or \eedd
parton cross sections generally only at the level of a percent (up to
10\,\% for $V_0$ and $V_1$). If all parton channels are summed up, the
impact on the total hadronic cross section is even smaller if the
heavier flavors are not excited by the contact interactions.
\\[2mm] 
(ii) The coupling for  single leptoquark production out of
the sea in positron scattering is large, of order $e$. However, for
such couplings the cross section \mbox{$\sigma(e^-p \longrightarrow
  {\raisebox{1ex}[-1ex]{\hspace{-0.75cm} $q$ \hspace{0.3cm}}} LQ)$}
would be two orders of magnitude larger than \mbox{$\sigma(e^+p
  \longrightarrow {\raisebox{1ex}[-1ex]{\hspace{-0.75cm} $\bar{q}$
      \hspace{0.3cm}}} LQ)$} and it seems unlikely that the large
number of events with $m_{LQ}\sim 200 $ GeV could have been missed in
the earlier electron-proton runs at HERA\footnote{Moreover, the low
  energy bounds restrict the couplings to $g_{L,R}\lsim 0.1$, see
  Table~\ref{tabprop}.}.  Nevertheless, Yukawa couplings of the order
$g_{L,R}\sim e/3$ do not seem to be ruled out yet
completely~\cite{oldhera}.  In such a scenario several types of
leptoquarks could be responsible for the observed events at HERA.  The
$F=2$ leptoquarks may lead to observable effects in hadron production
in \ee annihilation at LEP2, or at least more stringent bounds on the
Yukawa couplings can be established.
\\[2mm]
(iii) If the HERA events are interpreted as the signal of leptoquark
resonance production, they must originate from valence quarks, that is
from $e^+ u$ or $e^+ d$ fusion to $F=0$ states, and they must couple
chirally to an extremely good approximation in order to be consistent
with the existing bounds. Because of charge conservation, there is only
one possible process which can give rise to new events in
charged--current reactions,  $e^+ d \rightarrow \bar{\nu} u$,
and only in the presence of left-handed couplings. This implies that
an excess of events in the $CC$ channel, in addition to the $NC$ channel,  
will single out just two states:
the vector $LQ$s $V_0$ and $V_1$ carrying charge $\frac{2}{3}$. Thus,
even without switching from an $e^+$ to an $e^-$ beam, the $NC$ and
$CC$ searches combined are very selective in the $LQ$ quantum numbers.

\begin{figure}[htbp] 
\unitlength 1mm
\begin{picture}(162,155)
\put(-1.5,-1){
\epsfxsize=15.85cm
\epsfysize=14.6cm
\epsfbox{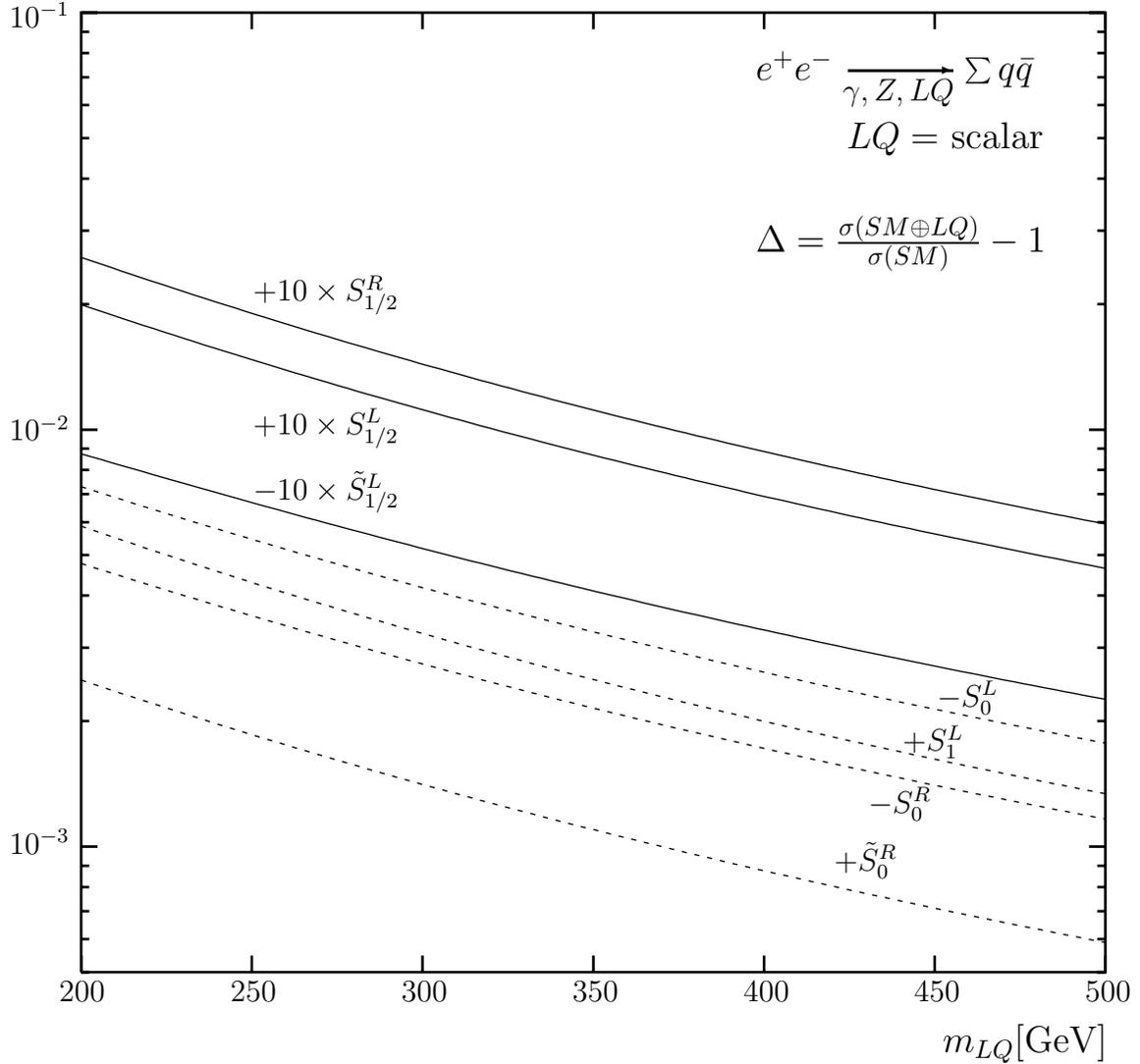}}
\setlength{\unitlength}{0.1bp}
\put(2985,3126){\makebox(0,0)[l]{\large $\Delta = \frac{\sigma(SM\oplus 
LQ)} {\sigma(SM)}-1$}}
\put(3333,3526){\makebox(0,0)[l]{\large $LQ = $ scalar}}
\put(2985,3758){\makebox(0,0)[l]{{\large $e^+e^-$} 
\raisebox{-1.5ex}{$\stackrel{}{\tiny \gamma, Z, LQ}$} 
{\large $\sum q\bar{q}$}}}
\put(3340,3790){\vector(1,0){400}}
\put(1056,2171){\makebox(0,0)[l]{$-10\times \tilde{S}_{1/2}^L$}}
\put(1056,2416){\makebox(0,0)[l]{$+10\times S_{1/2}^L$}}
\put(1056,2928){\makebox(0,0)[l]{$+10\times S_{1/2}^R$}}
\put(3287,756){\makebox(0,0)[l]{$+\tilde{S}_0^R$}}
\put(3418,972){\makebox(0,0)[l]{$-S_0^R$}}
\put(3550,1211){\makebox(0,0)[l]{$+S_1^L$}}
\put(3681,1382){\makebox(0,0)[l]{$-S_0^L$}}
\put(4018,51){\makebox(0,0){\large $m_{LQ}$[GeV]}}
\put(4337,251){\makebox(0,0){500}}
\put(3681,251){\makebox(0,0){450}}
\put(3025,251){\makebox(0,0){400}}
\put(2369,251){\makebox(0,0){350}}
\put(1712,251){\makebox(0,0){300}}
\put(1056,251){\makebox(0,0){250}}
\put(400,251){\makebox(0,0){200}}
\put(350,4004){\makebox(0,0)[r]{$10^{-1}$}}
\put(350,2416){\makebox(0,0)[r]{$10^{-2}$}}
\put(350,829){\makebox(0,0)[r]{$10^{-3}$}}
\end{picture}
\caption{\it Effect of $t/u$-channel exchange of scalar leptoquarks on the
  total hadronic cross section
as a function of $m_{LQ}$ for  $\protect\sqrt{s} = 192$ GeV.
The couplings have been fixed arbitrarily to  $(g_L,g_R) = (0.1,0)$ 
or $(0,0.1)$  indicated by $LQ^{L,R}$, respectively .}
\label{figxss}
\end{figure}
\begin{figure}[htbp] 
\unitlength 1mm
\begin{picture}(162,155)
\put(-1.5,-1){
\epsfxsize=15.85cm
\epsfysize=14.6cm
\epsfbox{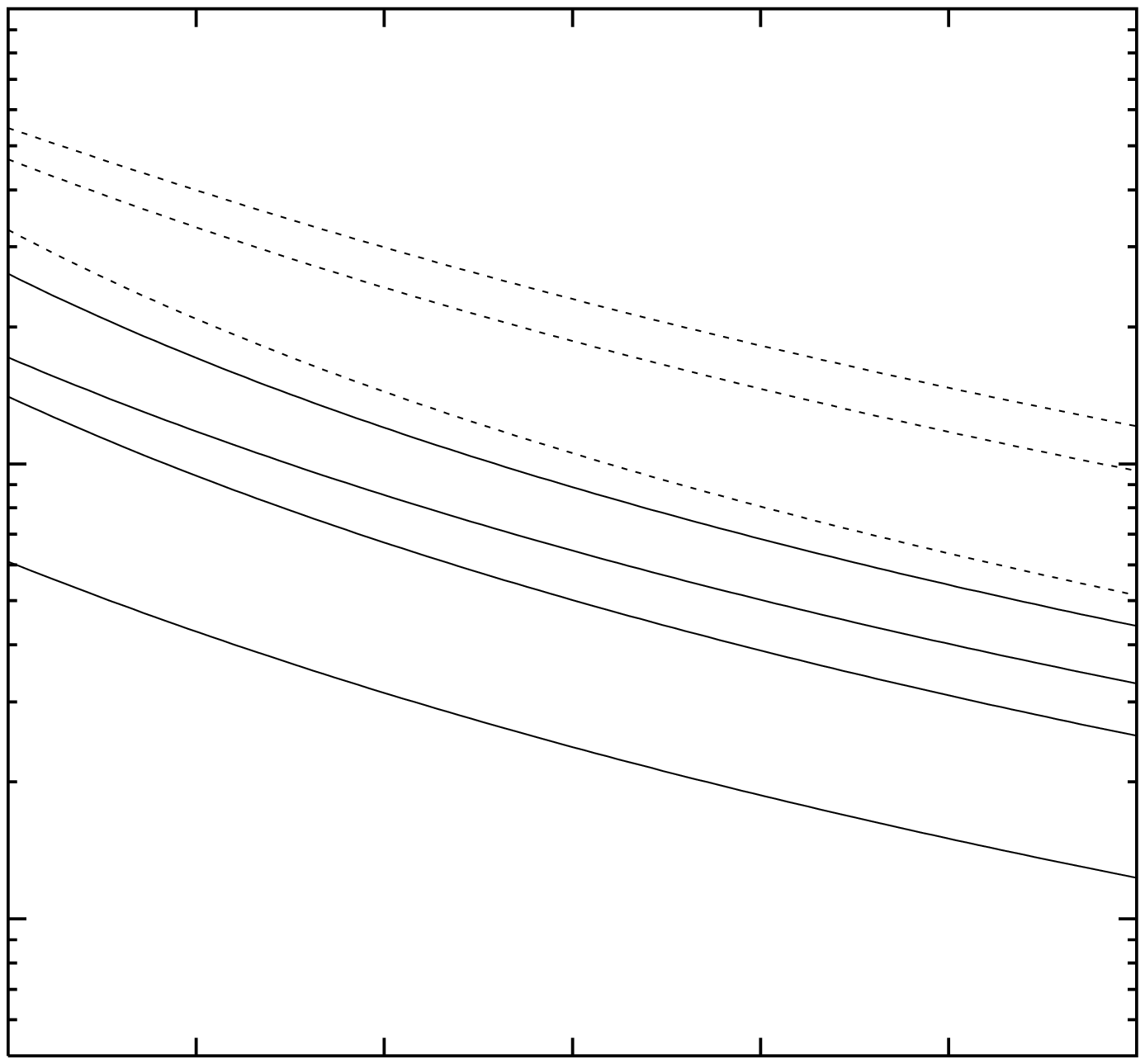}}
\setlength{\unitlength}{0.1bp}
\put(2985,3126){\makebox(0,0)[l]{\large $\Delta = \frac{\sigma(SM\oplus 
LQ)} {\sigma(SM)}-1$}}
\put(3333,3526){\makebox(0,0)[l]{\large $LQ = $ vector}}
\put(2985,3758){\makebox(0,0)[l]{{\large $e^+e^-$} \raisebox{-1.5ex}
{$\stackrel{}{\tiny \gamma, Z, LQ}$} {\large $\sum q\bar{q}$}}}
\put(3340,3790){\vector(1,0){400}}
\put(2631,2542){\makebox(0,0)[l]{$-10\times \tilde{V}_{1/2}^L$}}
\put(1712,2696){\makebox(0,0)[l]{$+10\times V_{1/2}^L$}}
\put(1712,3197){\makebox(0,0)[l]{$-10\times V_{1/2}^R$}}
\put(1056,1652){\makebox(0,0)[l]{$-V_0^R$}}
\put(1056,2218){\makebox(0,0)[l]{$+\tilde{V}_0^R$}}
\put(662,2696){\makebox(0,0)[l]{$-V_0^L$}}
\put(1056,2623){\makebox(0,0)[l]{$+V_1^L$}}
\put(4018,51){\makebox(0,0){\large $m_{LQ}$[GeV]}}
\put(4337,251){\makebox(0,0){500}}
\put(3681,251){\makebox(0,0){450}}
\put(3025,251){\makebox(0,0){400}}
\put(2369,251){\makebox(0,0){350}}
\put(1712,251){\makebox(0,0){300}}
\put(1056,251){\makebox(0,0){250}}
\put(400,251){\makebox(0,0){200}}
\put(350,4004){\makebox(0,0)[r]{$10^{-1}$}}
\put(350,2416){\makebox(0,0)[r]{$10^{-2}$}}
\put(350,829){\makebox(0,0)[r]{$10^{-3}$}}
\end{picture}
\caption{\it Effect of $t/u$-channel exchange of vector leptoquarks on the
  total hadronic cross section
as a function of $m_{LQ}$ for  $\protect\sqrt{s} = 192$ GeV.
The couplings have been fixed arbitrarily to  $(g_L,g_R) = (0.1,0)$ 
or $(0,0.1)$  indicated by $LQ^{L,R}$, respectively .}
\label{figxsv}
\end{figure}

However, recalling that no symmetry principles are known which give
rise to relations among Yukawa scalar-fermion couplings -- even within
isomultiplets they differ by nearly two orders of magnitude in the
Higgs sector -- there could be additional leptoquarks at higher masses
and with larger couplings that cannot be observed as resonance states
in the HERA experiments. We have therefore studied the sensitivity of
the total hadronic \ee cross section to the entire ensemble of scalar
and vector leptoquarks listed in Table~\ref{tabprop}. The 
result\footnote{Numerical cross checks have been performed with the help 
of CompHEP \cite{comphep} adapted to leptoquark processes.} is
illustrated in Figs.\ \ref{figxss} and \ref{figxsv} for the parameter
set $(g_L,g_R)=(0.1,0)$ or $(0,0.1)$ at $\sqrt{s}=192$ GeV 
as a function of the $LQ$
mass.  Since the couplings are arbitrary parameters, only the relative
size of the curves is relevant. For small enough couplings and large
enough masses the curves scale in $g^2_{L,R}/m_{LQ}^2$. We observe
both constructive and destructive interference effects, depending on
the type of quarks in the final state.  The impact of $I=0$, 1
leptoquarks on the hadronic cross section is larger than the impact of
$I=1/2$ leptoquarks.

For large masses, the exchange of leptoquarks can be described by contact
interactions. Depending on the type of leptoquark, different helicity
combinations of lepton and quark currents are affected in either \uu
or \dd final states. The sign of the contact interactions depends on
the fermion number as shown in the previous section. Potentially large
effects can be expected for \ee annihilation to hadrons. This is
exemplified for a series of $\Lambda$ values in Table~\ref{tabcon2}. 
The symbols $LL$ etc.\ denote the helicity of the lepton current
followed by the helicity of the quark current. The effect of the 
contact interaction ($CI$) is shown for the
total \ee hadronic cross section at LEP192 if only one of the 
$u$-type or $d$-type quarks is involved in the contact interactions. 
Present analyses of hadron production at LEP set limits to $\Lambda$ 
already at the level of about 1.5  to 2.5 TeV~\cite{opal}.

{\small
\begin{table}[ht]
\begin{center}
\begin{tabular}{|c||c|c|c||c|c|c||}
\hline
\rule{0mm}{5mm} 
 $ik$     
&\multicolumn{3}{c||}{ \uu final state}  
&\multicolumn{3}{c||}{ \dd final state}\\[1mm]
\cline{2-7} \rule{0mm}{5mm} 
$\Lambda[GeV]$
& $1.5$ & $2.5$ & $3.5$ 
& $1.5$ & $2.5$ & $3.5$ 
\lin
\rule{0mm}{5mm}
&\multicolumn{6}{c||}{$\eta = +1$}\\[1mm]\hline
\rule{0mm}{5mm}
$LL$  & $-0.11$ & $-0.14$ & $-0.09$ & $ 0.92$ & $ 0.23$ & $ 0.10$ \\
$RR$  & $ 0.08$ & $-0.07$ & $-0.05$ & $ 0.63$ & $ 0.12$ & $ 0.05$ \\
$LR$  & $ 0.30$ & $ 0.01$ & $-0.01$ & $ 0.52$ & $ 0.08$ & $ 0.003$ \\
$RL$  & $ 0.41$ & $ 0.05$ & $ 0.01$ & $ 0.30$ & $ 0.004$ & $-0.001$
\\[1mm]\hline
\hline
\rule{0mm}{5mm}
&\multicolumn{6}{c||}{$\eta = -1$}\\[1mm]\hline
\rule{0mm}{5mm}
$LL$  & $ 0.99$ & $ 0.26$ & $ 0.12$ & $-0.03$ & $-0.11$ & $-0.07$ \\
$RR$  & $ 0.81$ & $ 0.19$ & $ 0.08$ & $ 0.26$ & $-0.008$ & $-0.02$ \\
$LR$  & $ 0.59$ & $ 0.11$ & $ 0.04$ & $ 0.37$ & $ 0.03$ & $ 0.002$ \\
$RL$  & $ 0.48$ & $ 0.07$ & $ 0.02$ & $ 0.59$ & $ 0.11$ & $ 0.04$
\\[1mm]\hline
\end{tabular}
\caption{\it The effect of contact interactions with different helicities
on the cross section for hadron production in \ee annihilation: 
$\Delta=\sigma(SM \oplus CI)/\sigma(SM)-1$.}
\label{tabcon2}
\end{center}
\end{table}}

\section {Squarks in $R$-parity Breaking SUSY Models}
\def\Rs{R \hspace{-0.35em}/\;}
In the minimal supersymmetric extension of the Standard Model, 
the only renormalizable, gauge invariant operator
that couples squarks to quarks and leptons is given by
\beq
W_{\Rs}=\lambda'_{ijk}L^i_LQ^j_L\bar{D}^k_R \label{superp}
\eeq
in the superpotential \cite{suprp}.  The indices $ijk$ are
generation indices in the  left-handed doublets of leptons ($L$) and quarks
($Q$), and right-handed singlets of down-type quarks ($D$).  This
interaction term violates global invariance of $R$-parity, defined as
$R=(-1)^{3B+L+2S}$ which is $+1$ for particles and $-1$ for
superpartners. This interaction has also been considered in the context 
of the Aleph 4-jet events in Ref.~\cite{aleph}.

Expanding the superfields in terms of matter fields, the interaction
Lagrangian can be written as
\beq {\cal L}_{\Rs}=\lambda'_{ijk}\left[
\ti{u}^j_L\bar{d}^k_Re^i_L
+\overline{\ti{d}}^k_R(\bar{e}^i_L)^cu^j_L
+\ti{e}^i_L\bar{d}^k_Lu^j_L 
-\ti{\nu}^i_L\bar{d}^k_Rd^j_L-\ti{d}^j_L\bar{d}^k_R\nu^i_L
-\overline{\ti{d}}^k_R(\bar{\nu}^i_L)^cd^j_L \right] + h.c.
\eeq
where $u^j$ and $d^k$ stand for $u$- and $d$-type quarks, respectively, 
and the superscript $(~)^c$ denotes charge conjugate spinors.
The first two terms with $i=1$  are of particular interest since
they allow for resonant squark production in $e^+p$  scattering
at HERA via the subprocesses 
\beq
e^+ + d^k_R & \ra & \ti{u}^j_L \qquad\qquad(\ti{u}^j=\ti{u},\ti{c},\ti{t})
 \label{charm}\\
e^+ + \bar{u}^j_L& \ra& {\overline{\ti{d}}^k_R} \label{bottom}
\qquad\qquad (\ti{d}^k=\ti{d},\ti{s},\ti{b}) \eeq
If more than one of the couplings
\begin{figure}[htb]
\centerline{\psfig{figure=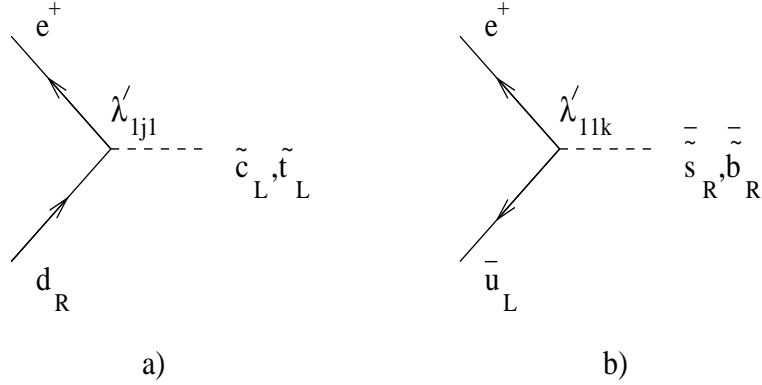,height=10cm,width=5cm,angle=-90}}
\caption{\it Squarks production mechanisms in $e^+p$ collisions.}
\label{figsusyn}
\end{figure}
$\lambda'_{1jk}$ is non-vanishing, 
strong limits on their product  are imposed by the absence 
of FCNC reactions~\cite{prod}.  Since
$\lambda'_{111}\lsim10^{-3}$ (for a mass $m=200$ GeV of 
supersymmetric partners
mediating neutrinoless double beta decay \cite{beta}), second or third
generation fermions must be coupled to electrons in order to account for
the rate at HERA.  Below we will consider the two
possible scenarios, shown in Fig.~\ref{figsusyn}, in which 
a) quarks, b) antiquarks from the proton participate in the production 
process. In particular we will discuss two cases  for $\lambda'_{1j1}\ne 
0$ or $\lambda'_{11k}\ne 0$ and their implications for  \ee 
annihilation to hadrons.

\vspace{2mm}
\noindent a) \underline{$\lambda'_{1j1}\ne 0$, $j=2,3$}
\begin{figure}[htbp]
\centerline{\psfig{figure=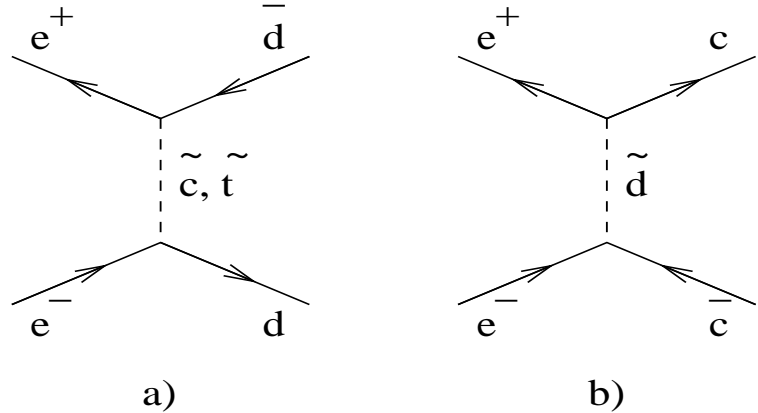,height=10cm,width=5.5cm,angle=-90}}
\caption{The scenario with $\lambda'_{1j1}\ne0$: 
$\ti{c}$ for $j=2$, $\ti{t}$ 
for $j=3$.}\label{figsusya}
\end{figure}

\begin{figure}[htbp] 
\unitlength 1mm
\begin{picture}(162,155)
\put(-1.5,-1){
\epsfxsize=15.85cm
\epsfysize=14.6cm
\epsfbox{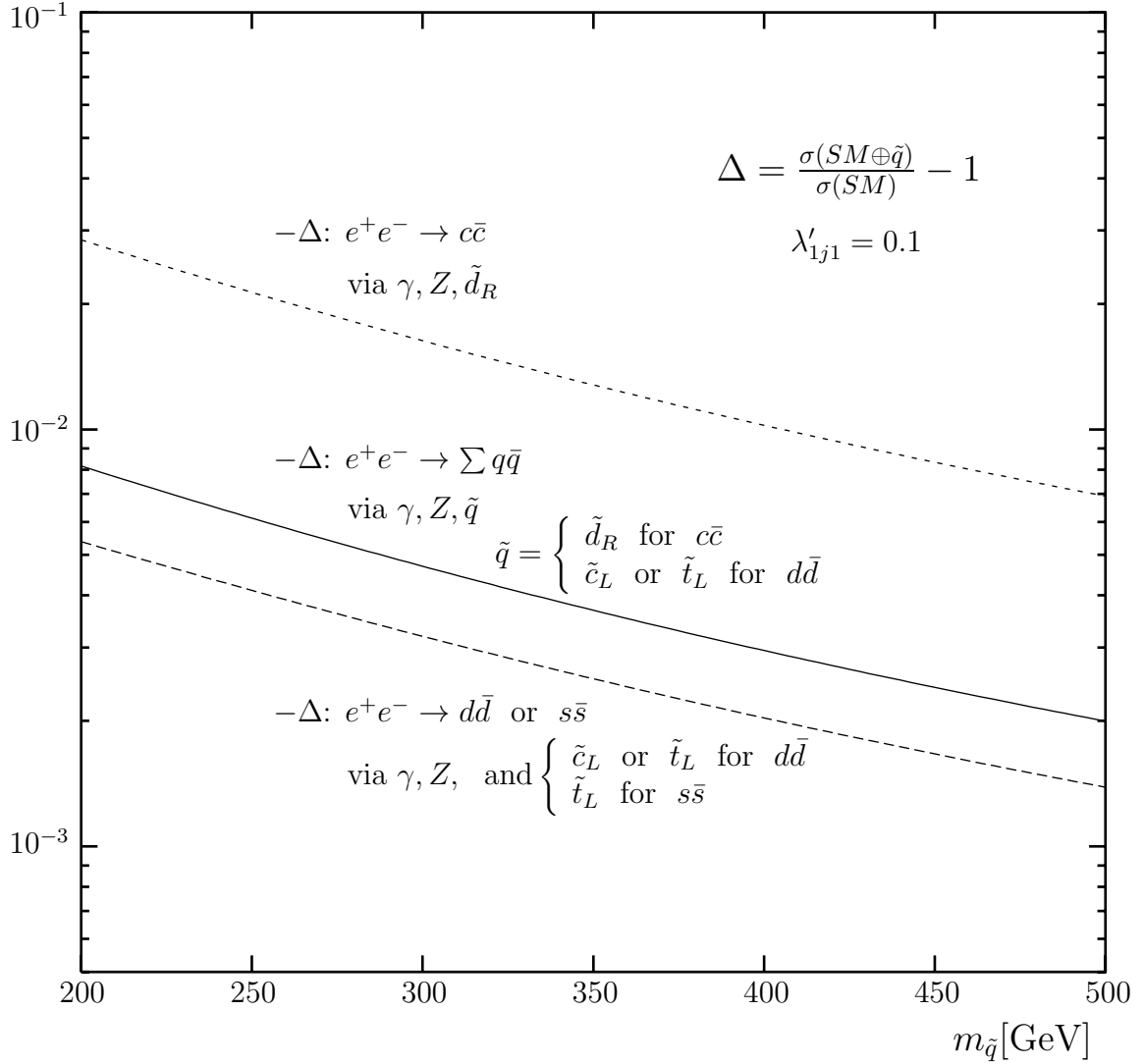}
}
\put(100,120){\makebox(0,0)[l]{\large $\Delta = \frac{\sigma(SM\oplus 
\tilde{q})}{\sigma(SM)}-1$}}
\put(110,110){\makebox(0,0)[l]{$\lambda'_{1j1}=0.1$}}
\put(70,68){\makebox(0,0)[l]{$\tilde{q}=\left\{\begin{array}{l}
 \tilde{d}_R~~{\rm for}~~c\bar{c}\\
 \tilde{c}_L~~{\rm or}~~\tilde{t}_L~~{\rm for}~~d\bar{d}
\end{array}\right.$}}
\put(50,74){\makebox(0,0)[l]{via $\gamma, Z, \tilde{q}$}}
\put(40,81){\makebox(0,0)[l]{$-\Delta$: $e^+e^- \rightarrow \sum q\bar{q}$}}
\put(50,105){\makebox(0,0)[l]{via $\gamma, Z, \tilde{d}_R$}}
\put(40,112){\makebox(0,0)[l]{$-\Delta$: $e^+e^- \rightarrow c\bar{c}$}}
\put(50,38){\makebox(0,0)[l]{via $\gamma, Z,~~{\rm and}
\left\{\begin{array}{l}
 \tilde{c}_L~~{\rm or}~~\tilde{t}_L~~{\rm for}~~d\bar{d}\\
 \tilde{t}_L~~{\rm for}~~s\bar{s} 
\end{array}\right.$}}
\put(40,47){\makebox(0,0)[l]{$-\Delta$: $e^+e^- \rightarrow 
d\bar{d}~~{\rm or}~~s\bar{s}$}}
\put(132,2){\makebox(0,0)[l]{\large $m_{\tilde{q}}$[GeV]}}
\setlength{\unitlength}{0.1bp}
\put(4337,251){\makebox(0,0){500}}
\put(3681,251){\makebox(0,0){450}}
\put(3025,251){\makebox(0,0){400}}
\put(2369,251){\makebox(0,0){350}}
\put(1712,251){\makebox(0,0){300}}
\put(1056,251){\makebox(0,0){250}}
\put(400,251){\makebox(0,0){200}}
\put(350,4004){\makebox(0,0)[r]{$10^{-1}$}}
\put(350,2416){\makebox(0,0)[r]{$10^{-2}$}}
\put(350,829){\makebox(0,0)[r]{$10^{-3}$}}
\end{picture}
\caption{\it Effect of $t/u$-channel exchange of squarks in 
  the supersymmetry scenario (a) on the total hadronic cross section,
  $\Delta = \sigma(SM \oplus \tilde{q}) / \sigma(SM) -1$, as a
  function of $m_{\tilde{q}}$ for $\lambda'_{1j1} = 0.1$, $j=2$ or $3$
  (or $\lambda'_{132}=0.1$ for $s\bar{s}$) and $\protect\sqrt{s} = 192$
  GeV.}
\label{figxsq}
\end{figure}

\noindent In this case,  $down$ quarks are 
involved\footnote{Another possibility is to consider $\lambda'_{132}\ne 0$
which would involve $strange$ quarks from the sea  in the 
production of $top$ squarks (Fig.~\ref{figsusyn}a).  
The qualitative analysis below applies to this case as well: 
$s\bar{s}$ production via $\ti{t}_L$ (see Fig.~\ref{figsusya}a and 
Fig.~\ref{figxsq}).}
via (\ref{charm}) in the production of heft-handed $charm$ ($j=2$) or
$top$ ($j=3$) squarks at HERA, Fig.~\ref{figsusyn}a.  The other
process (\ref{bottom}) is irrelevant since it would require $charm$ or
$top$ sea antiquarks in the proton.  To account for the
observed number of events at HERA, $\lambda_{1j1}$ must exceed 0.052
which is still within the limits derived from atomic parity violation
for a sufficiently heavy $\ti{d}_R$ 
\cite{Leurer,prod}. Notice that since $\ti{u}^j$ does not couple to neutrinos, 
the surplus of events in $CC$ reactions at HERA is not expected.
 In \eeqq the coupling $\lambda'_{1j1}$ leads to two
additional hadron channels, as shown in Fig.~\ref{figsusya}:
$e^+e^-\ra d\bar{d}$ via $charm$ or $top$ squark exchange in the
$t$-channel, and $e^+e^-\ra c\bar{c}$ via $down$ squark exchange in
the $u$-channel.  Since the left-handed $up$-type squarks couple in
the same way as the $\ti{S}_{1/2}$ leptoquark, and the right-handed
$down$ squark like the $S_0$ leptoquark with left-handed coupling
$g_L$, their contributions can easily be obtained from the formulae given
in the previous section.  The impact of these exchange mechanisms on
the total hadronic \ee cross section is shown in Fig.~\ref{figxsq}.
The impact of squark exchange on the single parton cross section $e^+e^-\ra
c\bar{c}$ is larger; the experimental analysis of  this process 
however requires the tagging of
$charm$ quarks in the final state.

\vspace{3mm}
\noindent b)\underline{ $\lambda'_{11k}\ne 0$, $k=2,3$}
\begin{figure}[htb]
\centerline{\psfig{figure=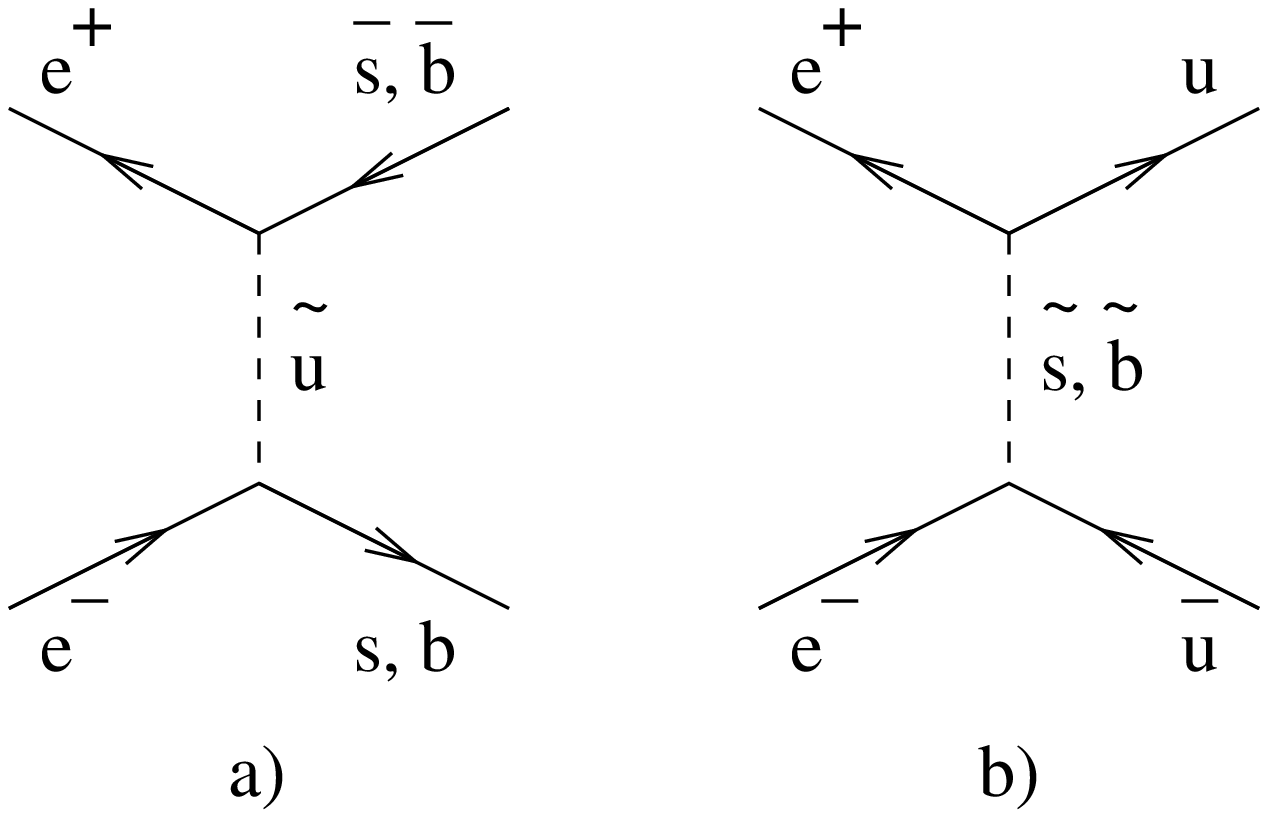,height=10cm,width=5.5cm,angle=-90}}
\caption{\it The scenario with $\lambda'_{11k}\ne0$: 
$\ti{s}$ for $k=2$, $\ti{b}$ 
for $k=3$.}\label{figsusyb}
\end{figure}

\noindent 
In this case, $up$ antiquarks of the sea are involved in the process
(\ref{bottom}) so that  $strange$ ($k=3$) or $bottom$ ($k=3$) 
antisquarks would be produced at HERA, Fig.~\ref{figsusyn}b.  
The $\ti{d}$ couples also to neutrinos, therefore similar events in $CC$ 
reactions could be expected. The coupling $\lambda'_{11k}$ would 
introduce two additional mechanisms in the \eeqq process, 
Fig.~\ref{figsusyb}: $e^+e^-\ra s\bar{s}$  or $e^+e^-\ra b\bar{b}$
via  $up$ squark exchange in the $t$-channel, and
$e^+e^-\ra u\bar{u}$ via  $strange$  or $bottom$ squark
exchange in the $u$-channel.  Since the parton densities of sea
quarks are much smaller than the densities of valence quarks, this
scenario would require a large coupling, $\lambda'_{11k}>0.3$.
Such a large value of $\lambda'_{11k}$ is however 
in conflict with the limit $\lambda'_{11k}<0.06$
derived from charged current universality \cite{CC} or from earlier
$e^-p$ data \cite{oldhera}.  This mechanism therefore cannot
explain the surplus of HERA data.

In contrast to genuine leptoquarks which decay solely to leptons and 
quarks, squarks can decay not only via
$R$-parity violating couplings but in general also  via a large 
number  of $R$-parity conserving modes:
$\ti{q}\ra q\chi$ with $\chi$ being either a neutralino or a chargino
state, cascading in a chain reaction down to ordinary particles.
The lower limits for the couplings $\lambda'$ inferred 
from the HERA events 
were based on  branching ratios of 100\,\% for the
$R$-parity violating decay modes to lepton plus quark jet. If the
branching ratios for $\ti{q}\ra q\chi$ decays are non-negligible, the
couplings $\lambda'$ would have to be larger correspondingly. 
This would increase the impact on \ee collisions. At the same time, new
types of events at HERA would be expected with multiple jet 
topologies and leptons from the $R$-parity breaking $\chi$ decays.

\newpage
\section{Summary}
The conclusions of our  analysis can be summarized in four
points.
\\[2mm]
(i) If the high $Q^2$, large $x$ events observed at HERA in
deep-inelastic positron-proton scattering are interpreted as the
direct production of a narrow leptoquark state, two cases must be
distinguished. For leptoquarks generated in collisions with valence
quarks, the Yukawa couplings are so small, $\sim e/10$, that the
contribution of the $t/u$-channel exchange of these leptoquarks
affects the production of hadrons in high-energy \ee annihilation only
at the level of  less than one percent. For leptoquarks generated
from antiquarks in the sea, the Yukawa couplings can still be
larger. However, the couplings presumably do not exceed the value
$e/3$ since the leptoquark states would have been observed otherwise
in the earlier electron-proton runs at HERA. In this second
case, the effects of leptoquark exchange may be accessible in \ee
annihilation at LEP2.
\\[2mm]
(ii) For masses above the range covered directly by HERA, the impact of
the leptoquark exchange on hadron production in \ee annihilation can be
significant for a wide  range of Yukawa couplings. The interactions are 
effectively described by contact terms, similar to contact interactions 
at HERA. The cross
sections have been presented for all standard leptoquark states.
\\[2mm]
(iii) In  $R$-parity breaking supersymmetric models, the HERA events
could be interpreted as the production of either left-handed
$charm$ or $top$ squarks. The observed number of events requires
$\lambda'_{1j1}>0.065$, the lower limit corresponding to a branching
ratio of $\sim 100$\% for the $R$-parity violating decays to lepton
and hadron jets. If the coupling is close to the lower limit, the
impact on the hadronic cross section in \ee annihilation is small,
$\sim 1$\%. If the coupling is stronger, the impact on  \ee
annihilation will be more pronounced. In this case, interesting 
multi-jet and lepton signatures due to the $R$-parity 
conserving decays $\ti{q}\ra q\chi$, followed by $R$-parity 
breaking $\chi$ decays, could occur  at HERA.
\\[2mm]
(iv) If the HERA events are not interpreted as the production of
narrow leptoquark resonances but are described globally by contact
interactions, the effective scale $\Lambda$ is predicted to be in
a range which is easily accessible at LEP2. However, if deviations from the 
prediction of the Standard Model for the cross section  of  
\ee annihilation to hadrons are not observed, contact interactions with
scales of order 2~TeV cannot account for the HERA
events.  This would restrict the interpretation of the events to the direct
production of narrow leptoquark-type  resonances.

\bigskip
\bigskip
\noindent
{\Large \bf Appendix}

If more than one leptoquark contributes to the same helicity amplitude, 
the expression in Eq.\ (\ref{total}) has to be supplemented by  
the interference terms between pairs of leptoquarks. For the \uu final 
states we find  
\beq 
\frac{N_c}{128\pi s}\Biggl[ 
+\frac{1}{2}g_{1L}^2 g^2_{3L} C_6(\tau_1,\tau_3)
-g_{1R}^2 g^2_{9R} C_7(\tau_9,\tau_1)
-2g_{10L}^2 g^2_{1L} C_7(\tau_{10},\tau_1) \\  
-2g_{10L}^2 g^2_{3L} C_7(\tau_{10},\tau_3)
-g_{6R}^2 g^2_{4R} C_7(\tau_4,\tau_6)
-g_{6L}^2 g^2_{5L} C_7(\tau_5,\tau_6) \Biggr]
\eeq
and for \dd final states
\beq
\frac{N_c}{128\pi s}\left[ 
+{2}g_{8L}^2 g^2_{10L} C_5(\tau_{10},\tau_8)
-g_{2R}^2 g^2_{8R} C_7(\tau_8,\tau_2)
-2g_{8L}^2 g^2_{3L} C_7(\tau_{8},\tau_3)\right. \\ \left.
-2g_{10L}^2 g^2_{3L} C_7(\tau_{10},\tau_3)
-g_{6R}^2 g^2_{4R} C_7(\tau_4,\tau_6)
-g_{7L}^2 g^2_{4L} C_7(\tau_4,\tau_7)\right]
\eeq
The numbering of
the leptoquark states and their couplings ($g_{iL}$ or $g_{iR}$) and 
masses ($\tau_i=m^2_i/s$) follows the
listing in Table~\ref{tabprop}, with $i=1,\dots,10$. 
For example $g_{6R}$ and $\tau_6$ refer
to the right-handed coupling and mass of the $S_{1/2}$ state.  The
functions $C_5$, $C_6$ and $C_7$ can be expressed in terms of $C_1$
and $C_2$ as follows:
\beq 
C_5(x,y)&=&\frac{C_1(x)-C_1(y)}{2y-2x}\non\\
C_6(x,y)&=&\frac{C_2(x)-C_2(y)}{2y-2x}\non\\
C_7(x,y)&=&\frac{C_1(x)+C_2(y)}{2+2x+2y}
\eeq

\bigskip
\noindent{\large \bf Acknowledgments:}\\[1mm]
 We have benefitted from 
discussions with E. Boos, W. Buchm\"uller, P. M\"attig and U.~Martyn.
\\[3mm]
\noindent NOTE: When finalizing the present paper we received copies
of Refs.~\cite{alt,dre}. The main conclusions are in mutual agreement.


\begin{thebibliography}{99}

\bibitem{sem} H1 Collab., C.\ Adloff et al., DESY 97-024 
      and {\it Z. Phys.\ C}, in press;\\ 
   ZEUS Collab., J.\ Breitweg et al., DESY 97-025 
      and {\it Z. Phys.\ C}, in press.

\bibitem{PS} J.C.\ Pati and A.\ Salam, \prd8,73, 1240, \ib10,74,
275.

\bibitem{GG} H. Georgi and S.L.\ Glashow, \prl32,74, 438;
   P.\ Langacker, \prep72,81, 185;
   J.\ L.\ Hewett and T.\ G.\ Rizzo, \prep183,89, 193.

\bibitem{composite}L. Abbott and E. Farhi, \plb101,81, 69, \npb189,81, 547;
  W.\ Buchm\"uller, Acta Phys.\ Austr.\ Suppl.\ XXVII (1985) 517; 
  B.~Schrempp and F.~Schrempp, \plb 153,85,101.

\bibitem{Rbroken} P. Fayet, \plb69,77,489; G. Farrar and P. Fayet, 
   \plb76,78,575; N. Sakai and T. Yanagida, \npb197,82,533.

\bibitem{suprp} 
  C.S. Aulah and R.N. Mohapatra, \plb119,82,316;
  F. Zwirner, \plb132,83,103;
  L.J. Hall and M. Suzuki, \npb231,84,419;
  S. Dawson, \npb261,85,297.

\bibitem{yanagida} H.\ Murayama and T.\ Yanagida, \mpl7,92, 147.

\bibitem{BRW} W.\ Buchm\"uller, R.\ R\"uckl and D.\ Wyler, \plb191,87,
442.


\bibitem{Aachen} M.\ Spira, T.\ K\"ohler,  
  Diploma Theses, RWTH Aachen (1989); 
A. Djouadi, T. K\"ohler, M. Spira and J. Tutas, \zpc46,90,679;
 B.~Schrempp,  Proceedings, {\it Physics at HERA}  
  (Hamburg 1991), eds.\ W.\ Buchm\"uller and G.\ Ingelman.  


\bibitem{BW} O.\ Shanker, \npb204,82, 253; 
   W.\ Buchm\"uller and D.\ Wyler, Phys.\ Lett.\ {\bf B177} (1986) 377.

\bibitem{Leurer} M.\ Leurer, \prd49,94, 333; \ib50,94, 536.

\bibitem{Davidson}S.\ Davidson, D.\ Bailey and B.\ Campbell,
   \zpc61,94, 613.

\bibitem{beta} M.\ Hirsch, H.V.\ Klapdor-Kleingrothaus and S.G.\
   Kovalenko, \prd53,96, 1329.

\bibitem{butt} J.L. Hewett, Proceedings, {\it 1990 Summer Study on High Energy 
  Physics -- Research Directions for the Decade} (Snowmass 1990), 
  ed.\ E. Berger (World Scientific, 1992); 
 J.~Butterworth and H.~Dreiner, \npb397,93,3.


\bibitem{pp} 
   J.L.~Hewett and S.~Pakvasa, \prd37,88, 3165;
   O.J.P. \'Eboli and A.V.~Olinto, \prd38,88, 3461;
   A.~Dobado, M.J.~Herrero and C.~Munoz, \plb207,88, 97;
   M.~de~Montigny and L.~Marleau, \prd41,90, 3523;
   B.~Dion, L.~Marleau and G.~Simon, LAVAL-PHY-96-16, hep-ph/9610397;
   T.G.~Rizzo, SLAC-PUB-7284, hep-ph/9609267;
   J.~Bl\"umlein, E.~Boos and A.~Kryukov, DESY 96-174, hep-ph/9610408;
   B.~Dion, L.~Marleau, G.~Simon and M.~de~Montigny, LAVAL-PHY-96-17,
   ALBERTA-THY-39-96, hep-ph/9701285.

\bibitem{ee} D.\ Schaile and P.M.~Zerwas, Proceedings, {\it Physics
   at Future Accelerators} (La Thuile, Geneva 1987), ed.\ J.H.~Mulvey,
   CERN 87-07;      
   N.D.\ Tracas and S.D.P. Vlassopulos, \plb220,89, 285;
   A.\ Djouadi, M.\ Spira and P.M.\ Zerwas, Proceedings, {\it $e^+e^-$
   Collisions at 500 GeV: The Physics Potential} (Munich, Annecy, Hamburg
   1991), ed.\ P.\ M.\ Zerwas, DESY 92-123B; 
   J.\ Bl\"umlein and R.\ R\"uckl, {\it ibid.};
   J.E.\ Cieza Montalvo and O.J.P.\ \'Eboli,  \prd47,93, 837; 
   J.\ Bl\"umlein and R.\ R\"uckl, \plb304,93, 337;
   D.\ Choudhury, \plb346,95, 291;
   J.\ Bl\"umlein, E.\ Boos and A.\ Kryukov, DESY 96-219, hep-ph/9610506;
   O.J.P.\ \'Eboli, M.C.\ Gonzalez-Garcia and J.K.\ Mizukoshi,
   FTUV/96-82, 
   IFIC/96-91, IFUSP-P 1251, hep-ph/9612254;
   M.A.\ Doncheski and S.\ Godfrey, hep-ph/9612385.

\bibitem{gamgam} J.E.\ Cieza Montalvo and O.J.P.\ \'Eboli,
see Ref.\ \cite{ee}.

\bibitem{BR} J.\ Bl\"umlein and R.\ R\"uckl, see Ref.\ \cite{ee}.

\bibitem{who} J.L.\ Hewett and S.\ Pakvasa, \plb227,89, 178;
   O.J.P.\ \'Eboli et al., \plb311,93, 147.  
   N.\ Nadeau and D.\ London, \prd47,93, 3742; G.~Belanger, 
   N.\ Nadeau and D.\ London, \prd49,94, 3140; 
   M.A.\ Doncheski and S.\ Godfrey, \prd49,94,6220, \ib51,95,1040.

\bibitem{pp2}   
   R.\ R\"uckl and P.M.\ Zerwas, Proceedings, {\it Physics
   at Future Accelerators} (La Thuile, Geneva 1987), ed.\ J.H.\ Mulvey,
   CERN 87-07; G. Bhattacharyya, D. Choudhury and K. 
   Sridhar, \plb349,95,118.


\bibitem{ee2} J.L.\ Hewett and T.G.\ Rizzo, \prd36,87, 3367;
   H.\ Dreiner et al., Mod.\ Phys.\ Lett.\ {\bf A3} (1988) 443;
   J.E.\ Cieza Montalvo and O.J.P.\ \'Eboli, \prd47,93, 837;
   D. Choudhury, \plb376,96,201;
   M.S.\ Berger, IUHET-343, hep-ph/9609517.

\bibitem{SZ} L.M.\ Sehgal and P.M.\ Zerwas, \npb183,81, 417.

\bibitem{debchou} D.\ Choudhury and S.\ Raychaudhuri, CERN-TH/97-26,
 hep-ph/9702392.

\bibitem{effl}E. Eichten, K. Lane and M. Peskin, \prl50,83,811; 
R. R\"uckl, \plb129,93,363, \npb234,84,91; see also the workshop report 
   P.~Haberl,  F.~Schrempp and H.-U. Martyn, in the Proceedings,
   {\it Physics at HERA}   
  (Hamburg, 1991), eds.\ W.\ Buchm\"uller and G.\ Ingelman.  


\bibitem{oldhera} H1 Collab., S.\ Aid et al., \plb369,96, 173;
   ZEUS Collab., M.\ Derrick et al., \plb306,93, 173.

\bibitem{comphep} P.A. Baikov et al., Proc. X Int.\ Workshop QFTHEP'95, 
hep-ph/9701412.

\bibitem{opal} P. M\"attig, Proceedings, {\it International Conference 
on High Energy Physics} (Warsaw 1996); OPAL physics note PN 280 and 
P.~M\"attig, private communication.

\bibitem{prod} K.\ Agashe and M.\ Graesser, \prd54,95, 4445;
  D.\ Choudhury and P.\ Roy, \plb378,96, 153.

\bibitem{aleph} M. Carena, G.F. Giudice, S. Lola and C.E.M. Wagner, 
CERN-TH/96-352, hep-ph/9612334.

\bibitem{CC} V.\ Barger, G.F.\ Giudice and T.\ Han, \prd40,89, 2987.
\bibitem{alt} G. Altarelli, J. Ellis, G.F. Giudice, S. Lola and 
M.L. Mangano, CERN-TH/97-40, hep-ph/9703276.
\bibitem{dre} H. Dreiner and P. Morawitz, hep-ph/9703279.

\end{thebibliography}
\end{document}